\documentclass[twocolumn,showpacs,preprintnumbers,amsmath,amssymb,prl]{revtex4}
\usepackage{graphicx}% Include figure files
\usepackage{dcolumn}% Align table columns on decimal point
\usepackage{bm}% bold math

\begin{document}

\preprint{}

\title{6-axis inertial sensor using cold-atom interferometry}
\author{B. Canuel}
\author{F. Leduc}
\author{D. Holleville}
\author{A. Gauguet}
\author{J. Fils}
\author{A. Virdis\footnote{Address: SAGEM D\'{e}fense et s\'{e}curit\'{e}, Groupe SAFRAN, URD11, 72-74 rue de la Tour Billy, BP 72 - 95101 Argenteuil Cedex, France}}
\author{A. Clairon}
\author{N. Dimarcq}
\author{Ch. J. Bord\'{e}}
\author{A. Landragin\footnote{Contact: arnaud.landragin@obspm.fr}}

\affiliation{LNE-SYRTE, CNRS UMR 8630, Observatoire de Paris, 61 avenue
de l'Observatoire, 75014 Paris, France}

\author{P. Bouyer}
 \affiliation{Laboratoire
Charles Fabry, CNRS UMR 8501, Centre Scientifique d'Orsay, $b\hat{a}t.$
503, BP 147, 91403 Orsay, France}

\date{\today}

\begin{abstract}
We have developed an atom interferometer providing a full inertial
base. This device uses two counter-propagating cold-atom clouds that
are launched in strongly curved parabolic trajectories. Three single
Raman beam pairs, pulsed in time, are successively applied in three
orthogonal directions leading to the measurement of the three axis
of rotation and acceleration. In this purpose, we introduce a new
atom gyroscope using a butterfly geometry. We discuss the present
sensitivity and the possible improvements.
\end{abstract}


\maketitle
% INTRO
Since its proof of principle in 1991 \cite{first,Chu}, atom
interferometry has demonstrated in particular great sensitivity to
accelerations \cite{gravi,gradio} and rotations \cite{Gustavson97}.
Among beautiful applications \cite{Berman97}, these experiments
offer attractive perspectives for application in inertial
navigation, geophysics or tests of fundamental physics
\cite{physfund}, where the ability of cold-atom interferometry to
give stable and accurate measurements can bring a real improvement
compared to standard technologies, as is already the case for atomic
clocks \cite{clock}. Nowadays, best performances are achieved by
interferometers using optical transitions \cite{Borde89,Borde91},
based on a sequence of three Raman pulses ($\pi /2-\pi-\pi /2$)
first introduced by M. Kasevitch and S. Chu \cite{Chu}. The pulses
couple the two hyperfine ground states ($|6S_{1/2}, F=3, m_{F}=0>$
and $|6S_{1/2}, F=4, m_{F}=0>$ in the case of Cesium atoms), which
split apart when using counter-propagating Raman lasers
\cite{raman}. The $\pi /2$ and $\pi$ pulses realize respectively the
beam-splitters and mirrors of the interferometer. This configuration
allows measurement of acceleration along the direction of
propagation of the Raman lasers. When the geometrical area included
in the interferometer is non zero, it also gives access to the
rotation around the axes perpendicular to the oriented area. Up to
now, atom interferometer have only been proven to be sensitive to a
single inertial quantity (e.g. Acceleration or rotation along one
single axis), although intrinsically sensitive to at least both
acceleration and rotation. In order to get full inertial monitoring,
all six axes (3 rotations and 3 accelerations) must be measured, as
needed for inertial navigation, geophysics measurements or some
tests of fundamental physics \cite{physfund}. In the past, this was
achieved by implementing multiple inertial sensors, as proposed in
\cite{physfund}, and the ability of using a single "proof mass" for
measuring all inertial axis has not yet been achieved. This
represents a real challenge for inertial measurement such as the
possibility of monitoring gravity and the 3 components of the earth
rotation at the same position.

\begin{figure}[htb]
\includegraphics[width=8cm]{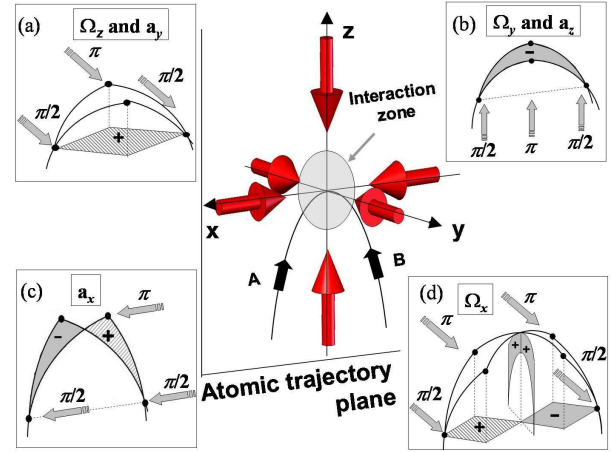}
\caption{6-axis inertial sensor principle. The atomic clouds are
launched on a parabolic trajectory, and interact with the Raman
lasers at the top. The four configurations (a)-(d) give access to
the 3 rotations and the 3 accelerations. In the three pulses
configuration, the Raman beams direction can be horizontal or
vertical, creating the interferometer in a horizontal (a) or
vertical (b)(c) plane. With a butterfly 4-pulse sequence of
horizontal beams (d), the rotation $\Omega_{x}$ can be measured. }
\label{fig:setup}
\end{figure}

In this letter we describe a new setup which is sensitive along six
axis of inertia. The two key features of our setup are the use of a
single Raman beam pair pulsed in time and the choice of a strongly
curved parabolic trajectory. This allows successive use of three
configurations of Raman lasers that interact with two
counter-propagating atomic clouds, giving access to all components
of rotation and acceleration. For one of theses components, we use a
new butterfly configuration based on a four-pulse sequence ($\pi
/2-\pi-\pi-\pi /2$). In addition, we introduce an original Raman
configuration to reduce the systematic effect introduced by
wavefront distortions.

%% GEOMETRIE DES LASERS (GENERAL)
In our experiment, about $10^{7}$ Cesium atoms are trapped from a
vapour in magneto-optical traps during 125 ms, and cooled down to 3
$\mu$K. The Cesium clouds are launched along parabolic trajectories
using moving molasses at 2.4 m.s$^{-1}$, with an angle of 8 degrees
with respect to the vertical direction. Then the atoms are prepared
in the state $|6S_{1/2}, F=3, m_{F}=0>$ before entering the
interferometer zone at the top of their trajectory, where they
interact with the Raman lasers. In the following \textbf{k} denotes
the effective wave vector of the Raman transition, and $\varphi_{l}$
the difference of phase between the two lasers. The interrogation
sequence is achieved with a single pair of Raman beams covering the
entire interrogation zone. The beams are switched on during 20
$\mathrm{\mu}$s to realize the Raman pulses, which provides an easy
way to change the pulse sequence. The atomic velocity and the Raman
beam size, 30 mm diameter (1/e$^{2}$), set the maximum interrogation
time to 80 ms. At the exit of the interferometer, the transition
probability depends on the inertial forces through the phase
difference accumulated between the two arms of the interferometer
\cite{Borde04}. Raman transitions enable detection of the internal
states of the atoms by fluorescence imaging.

We now present the description of the 6 axis inertial sensor
principle. The direction of sensitivity of the setup is defined by
the direction of the Raman interrogation laser with respect to the
atomic trajectory. As illustrated in Fig. \ref{fig:setup}, with a
classical three pulses sequence ($\pi /2-\pi-\pi /2$), a sensitivity
to vertical rotation $\Omega_{z}$ and to horizontal acceleration
$a_{y}$ is achieved by placing the Raman lasers horizontal and
perpendicular to the atomic trajectory \cite{Gustavson97}(Fig.
\ref{fig:setup}(a)). The same sequence, using vertical lasers, leads
to the measurement of horizontal rotation $\Omega_{y}$ and vertical
acceleration $a_{z}$ (Fig. \ref{fig:setup}(b)). Thanks to our
specific setup, we also have access to the other components of
acceleration and rotation which lie along the horizontal direction
of propagation of the atoms (x axis). The use of cold atoms in
strongly curved trajectories allows to point the Raman lasers along
the x direction, offering a sensitivity to acceleration $a_{x}$ and
no sensitivity to rotation (Fig. \ref{fig:setup}(c)). We also have
an easy access to the horizontal rotation $\Omega_{x}$ by changing
the pulse sequence to 4 pulses: $\pi /2-\pi-\pi-\pi /2$ (Fig.
\ref{fig:setup}(d)). We detail in the following the two
 configurations: the classical three pulses sequence (a) and our new butterfly four-pulse sequence (d).

The first pulse sequence that we study here is a standard three
pulses ($\pi /2-\pi-\pi /2$). The phase shift depends on the
acceleration \textbf{a} and on the rotation rate $\mathbf{\Omega}$
through \cite{Borde91}:

\begin{equation} \Delta\Phi=\textbf{k}(
\textbf{a}-2 ( \mathbf{\Omega} \times \textbf{v}))\mathrm{T^{2}}.
\label{equ1}
\end{equation}

The scale factor depends only on \textbf{k}, 2T the total
interrogation time and \textbf{v} the mean velocity in the
laboratory frame, which are well controlled. In the following
\textbf{k} is horizontal and along the y axis, as we see in Fig.
\ref{fig:setup}(a). The surface delimited by the two arms of the
interferometer is curved and the projection of the oriented area on
the two vertical planes cancels out. Therefore it gives access to
accelerations along this direction and to rotations around the
vertical axis. To discriminate between acceleration and rotation, we
use two counter-propagating Cesium atomic clouds leading to an
opposite velocity in eq. \ref{equ1} \cite{gyro}.

\begin{figure}[htb]
%\begin{center}
\includegraphics[width=5cm]{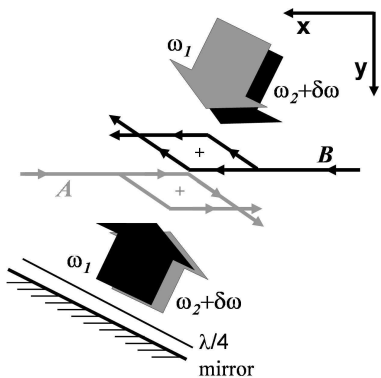}
%\end{center}
\caption{The orthogonally polarized copropagating Raman beams are
tilted with respect to the atom trajectories. They are
retroreflected by a mirror through a quater waveplate so that the
atoms interact with counterpropagating beams at frequency
$\mathrm{\omega}_{1}$ and
$\mathrm{\omega_{2}}+\mathrm{\delta\omega}$ with
$\mathrm{\omega_{1}-\omega_{2}}\approx$ 9.2 GHz. The detuning
$\delta\omega$ compensates for the Doppler shift so that each of the
two counterpropagating atom clouds can interact with only one pair
of beams. Interferometer areas are shown in the case of a
three-pulse interferometer.} \label{fig:mirror}
\end{figure}

%% Detail Laser
 In our setup, we have developed
a new method to reduce the variations of the local-wave vector
\textbf{k}, which induce perturbations that can be read as inertial
phase shifts \cite{Fils05}. In this method the Raman beams propagate
in the same optical system with orthogonal circular polarizations,
pass through the atomic trajectories and are retroreflected through
a quarter-wave plate\cite{Patent1}. In this case, the aberrations
are common and compensated most of the time: until the lasers cross
the atoms. With circular polarizations, the atoms can experience two
diffraction processes with opposite \textbf{k} vectors. In order to
select a single diffraction process, we tilt the laser beams by
6$^{o}$ in the horizontal plane (Fig.\ref{fig:mirror}), and
compensate the Doppler effect by an additional frequency difference
between the Raman lasers \cite{Doppler}. Since the two atom clouds
are counterpropagating, their Doppler detunings are opposite, which
means that each atomic cloud is resonant with a different Raman
pair, and this results in an opposite effective wave-vector for the
two interferometers. Therefore, the rotation and the acceleration
parts are respectively obtained by the sum and the difference of the
phases measured by the two interferometers (A and B).

\begin{figure}[htb]
%\begin{center}
\includegraphics[width=6cm]{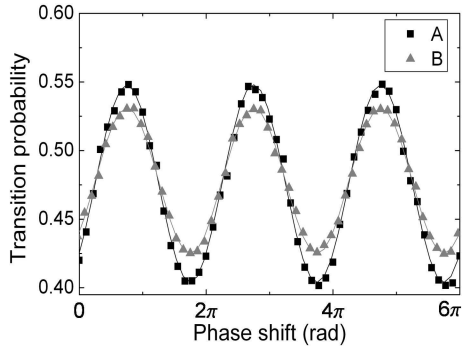}
%\end{center}
 \caption{Fringes obtained with the two interferometers A and B, for an interrogation time of 2T=60 ms.} \label{fig:franges3pulses}
\end{figure}

We show in Fig. \ref{fig:franges3pulses} the scan of the fringes of
both interferometers by changing the phase $\varphi_{l}$ between the
first and the second Raman pulse. With our interrogation time of
2T=60 ms, the fringe contrasts are respectively 14.4\% and 10.6\%
for A and B. The low contrast values can be explained by the sizes
of the clouds after ballistic expansion (3.3 mm rms radius) and by
the Gaussian intensity profile of the laser beams. In addition,
mismatch between the trajectories A and B requires a compromise for
the diffraction efficiency that leads to a reduction of the contrast
by a factor of about two.

\begin{figure}[htb]
%\begin{center}
\includegraphics[width=7cm]{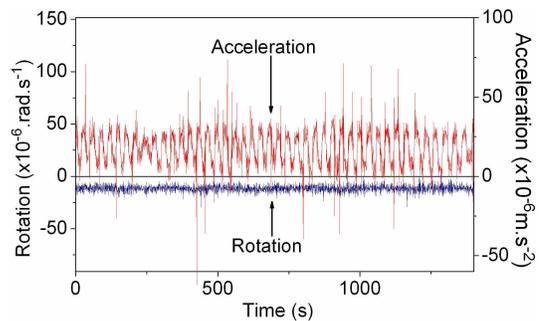}
%\end{center}
 \caption{Acceleration and rotation signals extracted from the half sum and half difference
of the phase shifts of interferometers A and B. DC offset on the
acceleration signal is due to the residual contribution of
gravitational acceleration. The acceleration dispersion
 on the acceleration signal comes from oscillations of the isolation platform.}
\label{fig:temp3pulses}
\end{figure}

To reach the maximum sensitivity to inertial forces, we operate the
interferometer on the side of a fringe. To realize this condition
for the two interferometers together, we align the Raman laser in
the horizontal plane and compensate the rotational phase with an
appropriate change of $\varphi_{l}$. In addition, by using two
different values of $\varphi_{l}$, the interferometers can sit
alternately on each side of a fringe \cite{clock}, which allows
rejection of long-term drifts of the contrast and of the offset of
the fringe patterns. Fig. \ref{fig:temp3pulses} shows the time
recordings of vertical rotation $\Omega_{z}$ and horizontal
acceleration $a_{y}$ extracted from the half sum and half difference
of the two interferometers' phase shifts.

\begin{figure}[htb]
%\begin{center}
\includegraphics[width=7cm]{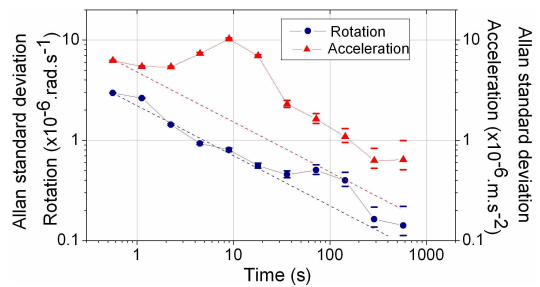}
%\end{center}
\caption{Allan standard deviations of acceleration (triangles)
and rotation (circles) measurements. Dashed lines corresponds
to the -1/2 slope expected for a white noise. The peak near 10 s
averaging time on the acceleration signal is due to the residual
oscillations of the isolation platform tilt.} \label{fig:allan}
\end{figure}

These results were obtained using an isolation platform (nano-K
350BM-1) to reduce the level of vibration in order to reach the
maximum sensitivity \cite{plateform}. However, this system
introduces long term tilt fluctuations which yields some
acceleration fluctuations through the projection of \textbf{g} on
the direction of \textbf{k}. To limit this effect, we have developed
a servo-lock of the platform tilt. The residual oscillation of this
system at 0.03 Hz can be identified on the acceleration signal.
Since this oscillation completely disappears on the rotation signal,
it gives a clear validation of the discrimination concept. We
estimate the performances of our setup from the Allan standard
deviation of these measurements. The signal-to-noise ratio from shot
to shot (0.56 s) is 12 for the acceleration and 39 for the rotation
leading to a respective sensitivity of $4.7\times10^{-6}$ m.s$^{-2}$
and $2.2\times10^{-6}$ rad.s$^{-1}$ for one second averaging time.
For both measurements, the Allan standard deviation (Fig.
\ref{fig:allan}) approaches the typical white noise behaviour for
long integration times. The sensitivity reaches $6.4\times10^{-7}$
m.s$^{-2}$ for acceleration and $1.4\times10^{-7}$ rad.s$^{-1}$ for
rotation after 10 min of averaging time.

We have performed the measurement of the Earth's rotation rate with
our cold atom interferometer: $5.50\pm0.05\times10^{-5}$
rad.s$^{-1}$, in which the error bar corresponds to statistical
uncertainty. This measured value for the projection along the
vertical axis was found in good agreement with the expected value at
Paris latitude ($\lambda=48^{o}50'08"$): $5.49\times10^{-5}$
rad.s$^{-1}$.

We now turn to the butterfly configuration (Fig.
\ref{fig:setup}(d)), which was first proposed to measure the gravity
gradient \cite{Gustavson-th}. It can be used to measure rotations
with the same Raman beams as in the previous configuration (y axis)
but in a direction (x axis) that cannot be achieved with a standard
3 pulses sequence. Four pulses, $\pi/2-\pi-\pi-\pi/2$, are used,
separated by times T/2-T-T/2 respectively. The atomic paths cross
each other leading to a twisted interferometer. The horizontal
projection of the oriented area cancels out so that the
interferometer is insensitive to rotation around the z axis. In
contrast, the vertical projection now leads to a sensitivity to
rotation around the x axis:

\begin{equation}
\Delta\Phi=\frac{1}{2}(\textbf{k} \times (\textbf{g}+\textbf{a})).
\Omega \mathrm{T}^{3}.
\end{equation}

This sensitivity to rotation appears from a crossed term with
acceleration and is no longer dependent on the launching velocity.
This configuration is not sensitive to DC accelerations along the
direction of the Raman laser, but remains sensitive to fluctuations
of horizontal and vertical acceleration. With our isolation
platform, the remaining fluctuations are negligible compared to
$\textbf{g}$, which does not compromise the stability of the scaling
factor. The sensitivity to rotation is comparable with that of
configurations (a) and (b). With 2T=60 ms, this configuration leads
to a interferometer area reduced by a factor 4.5, but it scales with
T$^{3}$ and thus would present a higher sensitivity for longer
interrogation times.

\begin{figure}[htb]
%\begin{center}
\includegraphics[width=6cm]{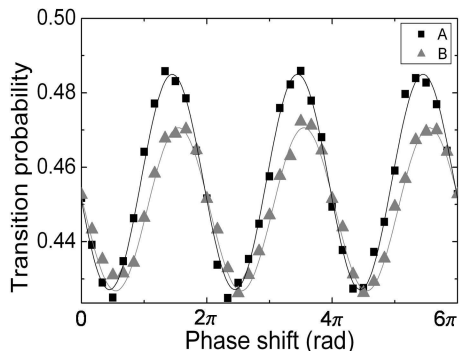}
%\end{center}
\caption{Fringes obtained with both interferometers A and B in the
4-pulses butterfly configuration for a total interrogation time of
2T=60 ms.} \label{fig:4pulsesfringes}
\end{figure}

The atomic fringe patterns are presented in Fig.
\ref{fig:4pulsesfringes} and show contrasts of 4.9\% and 4.2\% for
interferometer A and B respectively. By operating the interferometer
on the fringe side, as explained before, we obtain a signal-to-noise
ratio from shot to shot of 18 limited by the residual vibrations.
The sensitivity to rotation is equal to $2.2\times10^{-5}$
rad.s$^{-1}$ in 1 s, decreasing to $1.8\times10^{-6}$ rad.s$^{-1}$
after 280 s of averaging time.

To summarize, we have presented the ability to measure the 6
inertial axis with the same setup. This shows the advantage of using
cold atoms combined with a single laser beam pulse in the time
domain. A first measurement, in three pulses interferometer, has
demonstrated a sensitivity of $1.4\times10^{-7}$ rad.s$^{-1}$ to
rotation and $6.4\times10^{-7}$ m.s$^{-2}$ to acceleration in 10 min
averaging time. We have measured the Earth's rotation rate with an
accuracy of 1\%. Many improvements, such as the cold atom sources,
will allow to increase the sensitivity by a factor 50 on rotation
and 10 on acceleration \cite{Cheinet}.

In addition, we have demonstrated the butterfly configuration which
uses four pulses and which is sensitive to rotation around the axis
parallel to the direction of propagation of the atoms at the top of
their trajectory. This configuration is especially well adapted to
trajectories close to those of an atomic fountain, in which a single
source of atom is launched vertically. Since the interferometer area
scales with T$^{3}$, this opens the possibility of a cold atom
gyroscope reaching a sensitivity of $10^{-9}$ rad.s$^{-1}$ in one
second.

\begin{acknowledgments}
The authors would like to thank the D\'{e}l\'{e}guation
G\'{e}n\'{e}rale pour l'Armement, the Centre National d'Etudes
Spaciales, the SAGEM, the European Union (FINAQS) and the Ile de
France region (IFRAF) for their financial supports, Pierre Petit and
Christophe Salomon for their contributions to the early stage of the
experiment and Robert Nyman for careful reading of the manuscript.
\end{acknowledgments}


\begin{thebibliography}{30}
\bibitem{Doppler} With cold atoms, the Doppler width is small enough so that atoms can only feel the effect of a single Raman pair.
\bibitem{plateform} Without the platform, the fluctuations could induce phase shifts higher than 2$\pi$  \cite{gravi}, which would compromises the discrimination
process.
\end{thebibliography}
\end{document}